\documentstyle[12pt,amsfonts]{article}

\begin{document}

\title{The rovibrational kinetic energy for complexes of rigid molecules}
\author{Kevin A. Mitchell and Robert G. Littlejohn \\
Department of Physics, University of California, \\
Berkeley, California 94720 }
\date{\today}
\maketitle

\begin{abstract}
The rovibrational kinetic energy for an arbitrary number of rigid
molecules is computed.  The result has the same general form as the
kinetic energy in the molecular rovibrational Hamiltonian, although
certain quantities are augmented to account for the rotational energy
of the monomers.  No specific choices of internal coordinates or body
frame are made in order to accommodate the large variety of such
conventions.  However, special attention is paid to how key quantities
transform when these conventions are changed.  An example system is
explicitly analysed as an illustration of the formalism.
\end{abstract}

\section{Introduction}

The rovibrational kinetic energy of a molecular (rigid body) complex
has previously been computed explicitly for a number of specific
cases.  Examples include calculations for the atom-monomer system,
consisting of a point particle and a rigid body, by Brocks and van
Koeven\cite{Brocks88}, van der Avoird \cite{vanderAvoird93}, and
Makarewicz and Bauder \cite{Makarewicz95}; the molecular dimer,
consisting of two rigid bodies, by Brocks {\it et al.} \cite{Brocks83}
and van der Avoird \cite{vanderAvoird94}; the molecular trimer by
Xantheas and Sutcliffe\cite{Xantheas95} and van der Avoird, Olthof,
and Wormer \cite{vanderAvoird96}; as well as closely related systems
of single molecules with internal rotation \cite{Lister78,Gordy84}.
In the present article we derive the rovibrational kinetic energy of
an arbitrary molecular complex containing an arbitrary number of rigid
bodies.  That is, we express the kinetic energy in terms of the total
angular momentum of the complex and the momenta associated with the
internal degrees of freedom.  The term ``rovibrational'' is perhaps
misleading since it implies small amplitude vibrations of the complex,
an assumption we do not make.  A rovibrational kinetic energy for a
general molecular complex has been proposed earlier by Makarewicz and
Bauder \cite{Makarewicz95}; their analysis differs markedly from ours
in that they do not impose rigidity conditions on the monomers nor do
they distinguish the relative rotational motion of the monomers from
the internal vibrations of the monomers.

One method for deriving the rovibrational kinetic energy of a system
of rigid bodies would be to begin with the rovibrational kinetic
energy of a system of point particles and then impose rigidity
constraints within certain subsets of these particles.  A general
analysis of internal constraints on $n$ body systems has been given by
Menou and Chapuisat \cite{Menou93} and by Gatti {\it et al.}
\cite{Gatti97}.  These authors observe that such a formalism may be
applied to find the rovibrational kinetic energy of a rigid body
complex, though they do not derive such a kinetic energy.  The
constrained systems approach has the advantage of naturally allowing
the relaxation of the rigidity constraints to include small internal
vibrations of the monomers.  However, within a strictly rigid body,
the positions, masses, and velocities of the constituent particles are
irrelevant.  Rather, only the overall orientation, moment of inertia,
and angular velocity are of interest.  Therefore, in our derivation of
the rovibrational kinetic energy, we assume from the outset that the
bodies are rigid and use a kinetic energy consisting of only the
translational and rotational energies of the rigid bodies.  It should
be mentioned that whichever approach is followed, there exists an
ambiguous extrapotential term in the quantum kinetic energy.  The
origin of this ambiguity rests in the lack of knowledge about the
potential which confines the system to the manifold of rigid shapes.
See for example Kaplan, Maitra, and Heller \cite{Kaplan97}.  For
simplicity, we neglect any extrapotential terms arising from the
constraint process and adopt the standard form (translational plus
rotational energies) for the quantum kinetic energy of a rigid body.

Our approach is modeled on the derivation of
Ref.~\cite{Littlejohn97}, valid for clusters of point particles, but
is augmented to include the rotational kinetic energy of the monomers;
the current paper therefore generalises many results valid for point
particles to systems of extended rigid bodies.  Also, in deriving the
rovibrational kinetic energy, we make no specific choice of internal
coordinates or body frame.  This allows our results to be applicable
to the wide range of coordinate and frame conventions suitable to
different molecular complexes.  We discuss how the various
quantities appearing in the kinetic energy transform under changes of
internal coordinates and body frame.  We also present two
distinct decompositions of the kinetic energy into rotational and
vibrational (internal) contributions.  The first decomposition is
independent of the conventions for internal coordinates and body
frame.  The second decomposition has a form more common to
rovibrational Hamiltonians in the literature, though it is not
independent of coordinate and frame conventions.

The current paper has been  influenced by recent work
exhibiting the importance of gauge theory and geometric phase in the
study of rovibrational coupling
\cite{Littlejohn97,Guichardet84,Iwai87a,Iwai87b,Iwai88,Shapere89b,Montgomery96}.
In addition to rovibrational coupling, the geometric, or Berry's,
phase \cite{Shapere89a} has important applications in Born-Oppenheimer
theory\cite{Mead92}, optics\cite{Chiao86,Tomita86}, and guiding centre
motion\cite{Littlejohn88}, to name but a few examples.  The influence
of gauge theory on our analysis here is most readily evident in the
form of the kinetic energy Eq.~(\ref{r18}), which owing to the
appearance of a gauge potential is reminiscent of the kinetic energy
of a charged particle in a magnetic field.  Notions of gauge
invariance and covariance have also influenced our analysis of
transformation properties presented in Sect.~\ref{s5}.  Although our
development is influenced by the techniques and concepts of gauge theory
and geometric phase, this paper requires no specific background in either
field.

The article is structured as follows.  Section~\ref{s2} is devoted to
the classical kinetic energy.  The principal computation of the
article, culminating in the rovibrational kinetic energy
Eq.~(\ref{r18}), is contained in Sect.~\ref{s3}.  In Sect.~\ref{s4} we
show how the rovibrational kinetic energy of Eq.~(\ref{r18}) can be
placed in a different form, that of Eq.~(\ref{r21}), which is more
common in the rovibrational literature.  Sect.~\ref{s5} is a
systematic discussion of how important quantities transform under
changes in the internal coordinates and body frame.  Sect.~\ref{s1} is
an aside in which we discuss the form of quantum kinetic energy
operators in general.  In particular, Sect.~\ref{s7} discusses an
alternative to the usual Podolsky form of the quantum kinetic energy,
and Sect.~\ref{s6} illustrates how scaling the wave function
introduces an extrapotential term into the kinetic energy. In
Sect.~\ref{s8} we apply the results of Sect.~\ref{s1} to compute the
quantum kinetic energy of a molecular complex.  We work through a
concrete example in Sect.~\ref{s9} applying our formalism to the
system of a single monomer and an atom.  Sect.~\ref{s10} contains
conclusions.

\section{The classical kinetic energy of a molecular complex}

\label{s2}

\subsection{The principal derivation}

\label{s3}

In this paper, a molecular complex is modeled by a collection of $n$
rigid bodies or monomers.  No constraints are placed on the positions
or orientations of the monomers or on the symmetry of the moment of
inertia tensors.  In particular, we allow the monomers to be point
particles (atoms), collinear bodies, or noncollinear bodies.  However,
for mathematical simplicity, we initially assume that each monomer is
noncollinear.  Then, after deriving the rovibrational kinetic energy,
we comment on the straightforward generalisation of allowing complexes
containing collinear monomers and point particles.

We begin by introducing three classes of frames which are important in
our derivation.  The space frame (SF) is the inertial,
laboratory-fixed frame.  There are $n$ individual body frames (IBF),
one for each body; IBF $\alpha$, $\alpha = 1,\ldots,n$, is fixed to
rigid body $\alpha$ and rotates with the body.  Finally, there is a
single collective body frame (CBF), which is fixed to the complex as a
whole.  The CBF differs from the other frames in that it must be
specified for each shape or internal configuration of the complex.  In
general, this specification produces singularities in the CBF, a fact
which has been studied in three- and four-atom systems
\cite{Littlejohn98a,Littlejohn98b}.  There is no canonical way of
choosing the CBF, though there are several methods commonly used in
such Hamiltonians, such as fixing the CBF to the principal axes of the
complex or to one of the IBFs.  In this paper we will make no specific
choice of CBF.

We employ the following notation to denote the frame to which the
components of a vector are referred.  For an arbitrary vector ${\bf
v}$, an $s$ superscript, that is ${\bf v}^{s}$, indicates components
in the SF; an $i \alpha$ superscript indicates components in the IBF
of body $\alpha$; and a $c$ superscript indicates components in the
CBF.  (For notational simplicity, at a certain point we will drop the
$c$ superscript, leaving it understood thereafter that a vector
without a superscript is implicitly in the CBF.)  The components of
${\bf v}$ in the various frames are related by proper orthogonal $3
\times 3$ matrices, which we define by

\begin{eqnarray}
{\bf v}^{s} & = & {\sf R}{\bf v}^{c}, 
\label{r13} \\
{\bf v}^{s} & = & {\sf S}^{s}_\alpha {\bf v}^{i \alpha}, 
\label{r14} \\
{\bf v}^{c} & = & {\sf S}^{c}_\alpha {\bf v}^{i \alpha}.
\label{r15}
\end{eqnarray}
The matrix ${\sf S}^{s}_\alpha$ determines the orientation of body
$\alpha$, that is, the IBF of body $\alpha$, in the SF; the matrix
${\sf R}$ determines the orientation of the entire complex in the SF;
and the matrix ${\sf S}^{c}_\alpha$ determines the orientation of
body $\alpha$ in the CBF.

The configuration of the complex of $n$ rigid bodies is fully
specified by the centre of mass position of each body and the
orientation of each body.  To eliminate the overall translational
degrees of freedom, we fix the centre of mass of the entire complex at
the origin.  The centre of mass positions of the $n$ bodies are then
determined by $n-1$ Jacobi vectors \cite{Aquilanti86,Littlejohn97}
${\bf r}^{s}_{\alpha}$, $\alpha = 1,\ldots,n-1$.  The orientations of the
rigid bodies are specified by the $n$ matrices ${\sf S}^{s}_{\alpha}
\in SO(3)$, $\alpha = 1,\ldots,n$.  Taken together, ${\bf r}^s_\alpha$,
$\alpha = 1,\ldots, n-1$ and ${\sf S}_\alpha^s, \alpha = 1,\ldots, n$
specify a lab description of the configuration.  To shift to an
internal-external, or shape-orientation, description of the
configuration, we introduce $6n-6$ internal, or shape, coordinates
$q^\mu$, $\mu = 1,\ldots,6n-6$.  The $q^\mu$ may be separated into $3n -
6$ coordinates parametrising the distances between the rigid bodies
and $3n$ coordinates (Euler angles) parametrising the orientations of
the bodies in the CBF.  However, here we allow the $q^\mu$ to be
completely arbitrary, so long as they are invariant under rotations of
the complex.  The Jacobi vectors ${\bf r}^{c}_\alpha$, referred to the
collective body frame, and the matrices ${\sf S}^{c}_\alpha$ are both
functions of the internal coordinates $q^\mu$.  In practice, one may
define the CBF by specifying the functions ${\bf r}^{c}_\alpha(q)$ and
${\sf S}^{c}_\alpha(q)$, where $q$ without a superscript refers to the
collection of all coordinates $q^\mu$.  The orientation ${\sf R} \in
SO(3)$ (defined in Eq.~(\ref{r13})) of the complex together with the
shape $q^\mu$ fully determine the configuration in the SF, as may be
seen by the following equations

\begin{eqnarray}
{\bf r}^{s}_{\alpha} & = & {\sf R} {\bf r}^{c}_\alpha (q), 
\label{r2} \\
{\sf S}^{s}_{\alpha} & = & {\sf R} {\sf S}^{c}_\alpha (q).
\label{r6}
\end{eqnarray}
Equation~(\ref{r2}) is an application of Eq.~(\ref{r13}), and
Eq.~(\ref{r6}) is implied by Eqs.~(\ref{r13})--(\ref{r15}).  The above
equations relate the internal-external description of a configuration
(in terms of ${\sf R}$ and $q$) to the original description (in terms
of ${\bf r}^{s}_\alpha$ and ${\sf S}^{s}_\alpha$).

We assume that the masses of the $n$ bodies have been absorbed into
our definition of the Jacobi vectors.  Thus, the kinetic energy of the
complex, without the overall translational contribution, is

\begin{equation}
T = {1 \over 2} \sum_{\alpha = 1}^{n - 1} |\dot{{\bf r}}^{s}_{\alpha}|^2 +
{1\over 2} \sum_{\alpha = 1}^n
\mbox{\boldmath $\omega$}^{s}_{\alpha} \cdot {\sf M}^{s}_{\alpha} \mbox{\boldmath $\omega$}^{s}_{\alpha},
\label{r9}
\end{equation}
where the dot is used for time derivatives, ${\sf
M}^{s}_{\alpha}$ is the moment of inertia of body $\alpha$ in the
SF, and $\mbox{\boldmath $\omega$}^{s}_{\alpha}$ is the angular velocity
of body $\alpha$.  

In general, an angular velocity is a vector which measures the
rotation rate of one frame with respect to another frame.  The
components of this vector may be referred to either of these two
frames (or to an arbitrary third frame for that matter).  For example,
the angular velocity $\mbox{\boldmath $\omega$}^{s}_{\alpha}$ measures the
rotation rate of the IBF of body $\alpha$ with respect to the SF; its
components are referred to the SF.  

We introduce the notation ${\bf v} \times$ for the antisymmetric
matrix which maps a vector ${\bf u}$ into the vector ${\bf v} \times
{\bf u}$.  Then,

\begin{eqnarray}
\mbox{\boldmath $\omega$}^{s}_{\alpha} \times & = & \dot{\sf S}^{s}_{\alpha} {\sf S}^{s \; T}_{\alpha},
\label{r4} \\
\mbox{\boldmath $\omega$}^{i \alpha}_{\alpha} \times & = &  {\sf S}^{s \; T}_{\alpha}  \dot{\sf S}^{s}_{\alpha},
\label{r5}
\end{eqnarray}
where the $T$ superscript denotes the matrix transpose.  Note that
these two formulas are consistent with the change of basis relation
Eq.~(\ref{r14}), that is $\mbox{\boldmath $\omega$}^{s}_\alpha = {\sf
S}^{s}_\alpha \mbox{\boldmath $\omega$}^{i \alpha}_\alpha$, as may be seen from
the general relation

\begin{equation}
{\sf Q} ({\bf v} \times) {\sf Q}^T = ( {\sf Q} {\bf v} ) \times,
\label{r68}
\end{equation}
where the vector ${\bf v}$ is arbitrary and ${\sf Q} \in SO(3)$.

We proceed by expressing the kinetic energy Eq.~(\ref{r9}) in terms of
the internal velocities $\dot{q}^\mu$ and the total angular velocity
$\mbox{\boldmath $\omega$}^{c}$.  The total angular velocity measures the rotation
rate of the CBF with respect to the SF.  It has components in both the
SF and the CBF which are

\begin{eqnarray}
\mbox{\boldmath $\omega$}^{s} \times & = & \dot{\sf R} {\sf R}^T,
\label{r3} \\
\mbox{\boldmath $\omega$}^{c} \times & = & {\sf R}^T \dot{\sf R}. 
\label{r1}
\end{eqnarray}
These equations are analogous to Eqs.~(\ref{r4}) and (\ref{r5}).
Equation~(\ref{r1}) permits the time derivatives of Eqs.~(\ref{r2})
and (\ref{r6}) to be expressed as

\begin{eqnarray}
\dot{{\bf r}}^{s}_{\alpha} 
& = & \dot{\sf R} {\bf r}^{c}_{\alpha} + {\sf R} {\bf r}^{c}_{\alpha, \mu} \dot{q}^\mu
= {\sf R} ( \mbox{\boldmath $\omega$}^{c} \times {\bf r}^{c}_{\alpha} + {\bf r}^{c}_{\alpha, \mu} \dot{q}^\mu), 
\label{r12}\\
\dot{\sf S}^{s}_{\alpha} 
& = & \dot{\sf R} {\sf S}^{c}_{\alpha} + {\sf R} {\sf S}^{c}_{\alpha, \mu} \dot{q}^\mu
= {\sf R} \left[ \left(\mbox{\boldmath $\omega$}^{c} \times \right)  {\sf S}^{c}_{\alpha} 
+ {\sf S}^{c}_{\alpha, \mu} \dot{q}^\mu\right],
\label{r7}
\end{eqnarray}
where the ``$,\mu$'' subscript denotes the derivative with respect to
the coordinate $q^\mu$.  In the above equations, we have used the
convention, which we adopt for the remainder of the paper, that the
Greek indices $\mu$, $\nu$, $\ldots$ are implicitly summed from $1$ to $6n
- 6$ when repeated.  However, the Greek indices $\alpha$, $\beta$, $\ldots$ which label either the Jacobi vectors or monomers are summed
explicitly.  Combining Eqs.~(\ref{r4}) and (\ref{r7}) and using
Eq.~(\ref{r6}), we find

\begin{equation}
\mbox{\boldmath $\omega$}^{s}_{\alpha} \times 
= {\sf R}[ (\mbox{\boldmath $\omega$}^{c} \times) + {\sf S}^{c}_{\alpha, \mu} {\sf S}^{c \; T}_\alpha \dot{q}^\mu ] {\sf R}^T.
\label{r8}
\end{equation}
Since ${\sf S}^{c}_{\alpha, \mu} {\sf S}^{c \; T}_\alpha$ is
antisymmetric, we define a vector $\mbox{\boldmath $\tau$}_{\alpha \mu}$ such that

\begin{eqnarray}
\mbox{\boldmath $\tau$}^{c}_{\alpha \mu} \times 
& = & {\sf S}^{c}_{\alpha, \mu} {\sf S}^{c \; T}_\alpha,
\label{r16} \\
\mbox{\boldmath $\tau$}^{i \alpha}_{\alpha \mu} \times 
& = &  {\sf S}^{c \; T}_\alpha {\sf S}^{c}_{\alpha, \mu}.
\label{r17}
\end{eqnarray}
By comparison with Eqs.~(\ref{r4}) and (\ref{r5}), we see that
$\dot{q}^\mu \mbox{\boldmath $\tau$}^c_{\alpha \mu}$ is the angular velocity of the IBF $\alpha$
with respect to the CBF.  By inserting Eq.~(\ref{r16}) into
Eq.~(\ref{r8}) and using Eq.~(\ref{r68}), we find

\begin{equation}
\mbox{\boldmath $\omega$}^{s}_{\alpha} 
= {\sf R} (\mbox{\boldmath $\omega$}^{c} + \dot{q}^\mu \mbox{\boldmath $\tau$}^{c}_{\alpha \mu})
= \mbox{\boldmath $\omega$}^{s} + \dot{q}^\mu \mbox{\boldmath $\tau$}^{s}_{\alpha \mu}.
\label{r11}
\end{equation} 
The above equation expresses the angular velocity of the IBF of body
$\alpha$ with respect to the SF as the sum of the angular velocity of
the CBF with respect to the SF plus the angular velocity of the IBF of
body $\alpha$ with respect to the CBF.  However, this decomposition
has no inherent physical meaning since it depends on the convention
used to define the CBF.  By appropriately changing
this convention, either of these two terms could be made to vanish.

Using Eqs.~(\ref{r12}) and (\ref{r11}) in Eq.~(\ref{r9}), we write the
kinetic energy as

\begin{equation}
T = {1\over 2} \mbox{\boldmath $\omega$} \cdot {\sf M} \mbox{\boldmath $\omega$} 
+ {\bf a}_\mu \cdot \mbox{\boldmath $\omega$} \dot{q}^\mu
+ {1\over 2} h_{\mu \nu} \dot{q}^\mu \dot{q}^\nu,
\label{r10}
\end{equation} 
where

\begin{eqnarray}
{\sf M} & = & \sum_{\alpha = 1}^{n - 1} \left( r_\alpha^2 {\sf I} 
- {\bf r}_{\alpha} {\bf r}^T_{\alpha} \right)
+ \sum_{\alpha = 1}^n {\sf M}_{\alpha}
 =  \sum_{\alpha = 1}^{n - 1} \left( r_\alpha^2 {\sf I} 
- {\bf r}_{\alpha} {\bf r}^T_{\alpha} \right)
+ \sum_{\alpha = 1}^n {\sf S}_\alpha {\sf M}^{i \alpha}_{\alpha} {\sf S}^T_\alpha, 
\label{r43} \\
{\bf a}_\mu & = & \sum_{\alpha = 1}^{n - 1} {\bf r}_\alpha \times 
{\bf r}_{\alpha, \mu} +
\sum_{\alpha = 1}^n {\sf M}_{\alpha} \mbox{\boldmath $\tau$}_{\alpha \mu}
 =  \sum_{\alpha = 1}^{n - 1} {\bf r}_\alpha \times 
{\bf r}_{\alpha, \mu} +
\sum_{\alpha = 1}^n 
{\sf S}_\alpha {\sf M}^{i \alpha}_{\alpha} \mbox{\boldmath $\tau$}^{i \alpha}_{\alpha \mu}, 
\label{r44} \\
h_{\mu \nu} & = & \sum_{\alpha = 1}^{n - 1} {\bf r}_{\alpha, \mu} 
\cdot {\bf r}_{\alpha, \nu} +
\sum_{\alpha = 1}^n \mbox{\boldmath $\tau$}_{\alpha \mu} 
\cdot {\sf M}_\alpha \mbox{\boldmath $\tau$}_{\alpha \nu}
=  \sum_{\alpha = 1}^{n - 1} {\bf r}_{\alpha, \mu} 
\cdot {\bf r}_{\alpha, \nu} +
\sum_{\alpha = 1}^n \mbox{\boldmath $\tau$}^{i \alpha}_{\alpha \mu} 
\cdot {\sf M}^{i \alpha}_\alpha \mbox{\boldmath $\tau$}^{i \alpha}_{\alpha \nu}, 
\label{r45}
\end{eqnarray}
and where we henceforth suppress the $c$ superscript on vectors and
tensors referred to the CBF.  We have also used ${\sf I}$ for the
identity matrix and ${\sf M}_\alpha = {\sf R}^T {\sf M}^{s}_{\alpha}
{\sf R}$ and ${\sf M}^{i \alpha}_\alpha = {\sf S}^{s T}_\alpha {\sf
M}^{s}_{\alpha} {\sf S}^{s}_\alpha$ for the moment of inertia of body
$\alpha$ in the CBF and IBF $\alpha$ respectively.  In the
above equations, we present two expressions for each of the quantities
${\sf M}$, ${\bf a}_\mu$, and $h_{\mu \nu}$.  The first expression
involves quantities referred entirely to the CBF.  The second
expression is slightly more complex but has the advantage that all of
the dependence on ${\sf S}_\alpha(q)$ and ${\bf r}_\alpha(q)$ is shown
explicitly by ${\sf S}_\alpha$, ${\bf r}_\alpha$, ${\bf r}_{\alpha,
\mu}$, and $\mbox{\boldmath $\tau$}^{i \alpha}_{\alpha \mu}$.  Note that ${\sf
M}^{i \alpha}_\alpha$ is the moment of inertia of body $\alpha$ in its
own IBF and as such is a constant matrix independent of both the shape
and orientation.

By rearranging the terms in Eq.~(\ref{r10}), we put the kinetic energy
in the form

\begin{equation}
T = {1\over 2} (\mbox{\boldmath $\omega$} + {\bf A}_\mu \dot{q}^\mu) \cdot {\sf M} 
 (\mbox{\boldmath $\omega$} + {\bf A}_\nu \dot{q}^\nu)
+ {1\over 2} g_{\mu \nu} \dot{q}^\mu \dot{q}^\nu,
\end{equation} 
where

\begin{eqnarray}
{\bf A}_\mu & = & {\sf M}^{-1} {\bf a}_\mu, 
\label{r52}\\
g_{\mu \nu} & = & h_{\mu \nu} - {\bf a}_\mu \cdot {\sf M}^{-1} {\bf a}_\nu.
\label{r20}
\end{eqnarray}
Converting the velocities to momenta, we find

\begin{eqnarray}
{\bf J} & = & {\partial T \over \partial \mbox{\boldmath $\omega$}} = 
{\sf M}(\mbox{\boldmath $\omega$} + {\bf A}_\mu \dot{q}^\mu), \\
p_\mu & = & {\partial T \over \partial \dot{q^\mu}} =
g_{\mu \nu} \dot{q}^\nu + {\bf A}_\mu \cdot {\bf J}.
\end{eqnarray}
The vector ${\bf J}$ is the total angular momentum of the complex.  It
satisfies the usual body-referred ``anomalous'' commutation relations
$\{J_i, J_j\} = - \sum_k \epsilon_{ijk}J_k$, where $\epsilon_{ijk}$ is
the usual Levi-Civita symbol.  The classical kinetic energy is now
expressible in terms of momenta as

\begin{equation}
T = {1\over 2} {\bf J} \cdot {\sf M}^{-1}  {\bf J} +
{1\over 2} (p_\mu - {\bf A}_\mu \cdot {\bf J})  g^{\mu \nu} 
(p_\nu - {\bf A}_\nu \cdot {\bf J}),
\label{r18}
\end{equation}
where $g^{\mu \nu}$ is the inverse of $g_{\mu \nu}$.

Equation~(\ref{r18}) decomposes the kinetic energy into two terms.
The first term is the kinetic energy the complex would have if it were
a rigid body of fixed shape.  We regard this term as the rotational
kinetic energy of the complex.  The second term we regard as the
internal, or vibrational, kinetic energy.  Often in such
rotation-vibration decompositions, a different rotational term appears
which contains a modified moment of inertia tensor and a modified
angular momentum vector.  We will relate the above decomposition to
such alternative decompositions in the next section.  For now,
however, notice that the appearance of ${\bf A}_\mu$ in the internal
kinetic energy couples the internal degrees of freedom to the angular
momentum.  For this reason, we call ${\bf A}_\mu$ the Coriolis
potential.  Furthermore, we call $g_{\mu \nu}$ the internal metric
because it acts to square $p_\mu - {\bf A}_\mu \cdot {\bf J}$.  Note
that the internal kinetic energy has the same ``$|{\bf p} - e {\bf
A}|^2$'' form as the kinetic energy of a particle in a magnetic field,
where the role of the vector, or gauge, potential is played by the
Coriolis potential and the role of the electric charge is played by
the angular momentum.

The kinetic energy given in Ref.~\cite{Littlejohn97} for a collection
of point particles has exactly the same form as Eq.~(\ref{r18}).
However, for collections of point particles, the quantities defined in
Eqs.~(\ref{r43}) -- (\ref{r45}) do not contain the terms with ${\sf
M}^{i \alpha}_\alpha$.  The appearance of these terms and, of course, the
introduction of an extra $3n$ internal coordinates are the sole
modifications necessary to augment the kinetic energy of a system of
point particles to include the rotational kinetic energy of the
monomers.  The fundamental reason why the form of the kinetic energy
is the same for these two cases is the rotational symmetry of the
kinetic energy operator; Eq.~(\ref{r18}) is in fact a general result,
valid for any $SO(3)$ invariant metric.  We will discuss this matter
further in a future publication.

We comment now on how the preceding results are generalised to include
collinear monomers and point particles.  First, the number of
coordinates changes.  A noncollinear body requires three Euler angles
to fully specify its orientation, whereas a collinear body requires
only two spherical coordinates to specify its orientation (the
direction of its collinear axis).  A point particle, of course,
requires no orientational coordinates.  Therefore, instead of $6n-3$
coordinates as before, there are $3n + 3 n_n + 2 n_c - 3$ coordinates
parametrising the centre of mass system.  Here, $n = n_n + n_c + n_p$
is the total number of monomers, where $n_n$ is the number of
noncollinear monomers, $n_c$ is the number of collinear monomers,
and $n_p$ is the number of point particles.

We next describe the form of the rovibrational kinetic energy when
incorporating collinear bodies and point particles.  First, the
computation leading from Eq.~(\ref{r9}) to Eq.~(\ref{r18}) is
essentially unchanged by the inclusion of collinear bodies and point
particles, so long as one takes the moment of inertia tensor ${\sf
M}_\alpha$ of a point particle to be $0$.  The moment of inertia
tensor of a collinear body is explicitly ${\sf M}_\alpha = \kappa_\alpha
({\sf I} - {\bf n}_\alpha {\bf n}_\alpha^T)$, where ${\bf n}_\alpha$
is a unit vector pointing along the collinear axis and $\kappa_\alpha$ is
the single nonzero principal moment.  The quantity ${\bf n}_\alpha$ is
a more natural measure of the orientation of a collinear body than
${\sf S}_\alpha$, since ${\sf S}_\alpha$ overparametrises the
orientations.  We therefore rewrite Eqs.~(\ref{r43}) -- (\ref{r45}) in
the following form, more appropriate for a general complex,

\begin{eqnarray}
{\sf M} & = &  \sum_{\alpha = 1}^{n - 1} \left( r_\alpha^2 {\sf I} 
- {\bf r}_{\alpha} {\bf r}^T_{\alpha} \right)
+ \sum_{\alpha = 1}^{n_n} {\sf S}_\alpha {\sf M}^{i \alpha}_{\alpha} {\sf S}^T_\alpha
+  \sum_{\alpha = n_n + 1}^{n_n + n_c} \kappa_\alpha \left( {\sf I} 
- {\bf n}_{\alpha} {\bf n}^T_{\alpha} \right), 
\label{r63}\\
{\bf a}_\mu & = &  \sum_{\alpha = 1}^{n - 1} {\bf r}_\alpha \times 
{\bf r}_{\alpha, \mu} +
\sum_{\alpha = 1}^{n_n} 
{\sf S}_\alpha {\sf M}^{i \alpha}_{\alpha} \mbox{\boldmath $\tau$}^{i \alpha}_{\alpha \mu} + 
 \sum_{\alpha = n_n + 1}^{n_n + n_c} \kappa_\alpha {\bf n}_\alpha \times 
{\bf n}_{\alpha, \mu},  
\label{r67} \\
h_{\mu \nu} & = &  \sum_{\alpha = 1}^{n - 1} {\bf r}_{\alpha, \mu} 
\cdot {\bf r}_{\alpha, \nu} +
\sum_{\alpha = 1}^{n_n} \mbox{\boldmath $\tau$}^{i \alpha}_{\alpha \mu} 
\cdot {\sf M}^{i \alpha}_\alpha \mbox{\boldmath $\tau$}^{i \alpha}_{\alpha \nu} +
\sum_{\alpha = n_n+1}^{n_n + n_c} \kappa_\alpha {\bf n}_{\alpha, \mu} 
\cdot {\bf n}_{\alpha, \nu},
\label{r65}
\end{eqnarray}
where we have ordered the monomers with the noncollinear bodies first,
the collinear bodies second, and the point particles last.  We omit
the straightforward proof of these equations relying instead on the
following two observations.  First, the orientational contribution from
the point particles has dropped out since ${\sf M}_\alpha$ is $0$ for
such particles.  Next, the orientational contribution from a collinear
body is identical to the contribution of a single Jacobi vector.  This fact
is easily understood by modeling a collinear body as two point
particles connected by a Jacobi vector.  The above equations may be
combined with Eqs.~(\ref{r52}) and (\ref{r20}) to obtain the kinetic
energy Eq.~(\ref{r18}) of a general molecular complex.  The only note
of caution occurs if the entire complex should become collinear, in
which case ${\sf M}$ is not invertible and singularities may arise.

\subsection{An alternative form of the kinetic energy}

\label{s4}

We rearrange the kinetic energy Eq.~(\ref{r18}) to place it in a form
which is more common in the literature on rovibrational Hamiltonians.
We define the modified moment of inertia tensor $\tilde{\sf M}$ by

\begin{equation}
\tilde{\sf M} = {\sf M} - h^{\mu \nu} {\bf a}_\mu {\bf a}_\nu^T,
\label{r19}
\end{equation}
where $h^{\mu \nu}$ is the inverse matrix of $h_{\mu \nu}$, and the
 vector ${\bf K}$, often called the angular momentum of vibration, by

\begin{equation}
{\bf K} = h^{\mu \nu} {\bf a}_\mu p_\nu.
\label{r62}
\end{equation}
We use these definitions to place the kinetic energy in the form

\begin{equation}
T = {1\over 2} ({\bf J} - {\bf K}) \cdot \tilde{\sf M}^{-1} ({\bf J} - {\bf K}) 
+ {1\over 2} p_\mu  h^{\mu \nu} p_\nu.
\label{r21}
\end{equation}
The above equation may be verified by comparing the terms of order
$J^2$, $J^1$, and $J^0$ in Eqs.~(\ref{r21}) and (\ref{r18}).  The
equality of the respective terms is readily apparent from the
following identities

\begin{eqnarray}
g^{\mu \nu} {\sf M}^{-1} {\bf a}_\nu 
& = & h^{\mu \nu} \tilde{\sf M}^{-1} {\bf a}_\nu,
\label{r22} \\
g^{\mu \nu} 
& = & h^{\mu \nu} + h^{\mu \sigma}({\bf a}_\sigma \cdot \tilde{\sf M}^{-1} {\bf a}_\tau )h^{\tau \nu}, 
\label{r23} \\
\tilde{\sf M}^{-1} 
& = & {\sf M}^{-1} + g^{\mu \nu} {\bf A}_\mu {\bf A}_\nu^T,
\label{r24}
\end{eqnarray}
which follow from Eqs.~(\ref{r19}) and (\ref{r20}).

Equation (\ref{r21}) provides an alternative rovibrational
decomposition of the kinetic energy, in a form common in the
literature for rovibrational Hamiltonians.  For example, the
Wilson-Howard-Watson molecular Hamiltonian\cite{Kroto75,Papousek82} is
expressed in this manner using the Eckart conventions, for which
various simplifications occur.  

\subsection{Changing the internal coordinates and the collective body frame}

\label{s5}

When the conventions for the internal coordinates or the CBF are
changed, the quantities defined in this paper do not, in general,
remain invariant.  Instead, they transform via a precise set of rules.
Although the analysis of these rules is not critical to the logical
flow of this paper, we include an account of them for the following two reasons.  An
understanding of transformation rules facilitates conversion between
different sets of conventions.  This is important in actual problems,
where it is not uncommon to utilise more than one coordinate or frame
convention, especially for large amplitude motions.  Also, knowledge
of transformation properties leads to the definition and consideration
of quantities which transform in a simple manner (that is invariantly
or covariantly).  Such quantities often have special geometric or
physical significance.  The review of Littlejohn and Reinsch
\cite{Littlejohn97} contains an in depth discussion of transformation
properties specialised for systems of point particles.  Since these
results are essentially unchanged by the generalisation to include
rigid bodies, we simply summarise here the key results from
Ref.~\cite{Littlejohn97} and comment on how they are to be extended.

We first introduce the important concept of $q$-tensors.  We consider
a new set of coordinates ${q'}^\mu$ which are functions of the old
coordinates $q^\mu$.  A $q$-tensor transforms under such a change in
coordinates by contracting each lower Greek index $\mu$ with $\partial
q^\mu/ \partial {q'}^\nu$ and each upper Greek index $\mu$ with
$\partial {q'}^\nu/ \partial q^\mu$.  The rank of the tensor is the
total number of such indices, whether upper or lower.  For example,
$\mbox{\boldmath $\tau$}_{\alpha \mu}$ is a rank one $q$-tensor because it
transforms via

\begin{equation}
{\mbox{\boldmath $\tau$}'}_{\alpha \mu}
= {\partial {q}^\nu \over \partial {q'}^\mu} \mbox{\boldmath $\tau$}_{\alpha \nu},
\end{equation}
where ${\mbox{\boldmath $\tau$}'}_{\alpha \mu}$ is computed from Eq.~(\ref{r16})
using the new coordinates ${q'}^\mu$ and ${\mbox{\boldmath $\tau$}}_{\alpha \mu}$
is computed using the old coordinates ${q}^\mu$.  Other rank one
$q$-tensors are $\mbox{\boldmath $\tau$}^{i \alpha}_{\alpha \mu}$, ${\bf a}_\mu$,
${\bf A}_\mu$, $p_\mu$, and $\dot{q}^\mu$.  Rank zero $q$-tensors,
also called $q$-scalars, are invariant under coordinate
transformations.  They include ${\sf S}^s_\alpha$, ${\sf S}_\alpha$,
${\sf R}$, ${\bf r}^s_\alpha$, ${\bf n}^s_\alpha$,
$\mbox{\boldmath $\omega$}^s_\alpha$, $\mbox{\boldmath $\omega$}^s$, ${\sf M}^s_\alpha$, ${\sf
M}^s$, $\tilde{\sf M}^s$, $T$, ${\bf J}^s$, and ${\bf K}^s$, as well
as these same quantities referred (where appropriate) to the CBF or
IBF $\alpha$.  Rank two $q$-tensors include $h_{\mu \nu}$, $h^{\mu
\nu}$, $g_{\mu \nu}$, and $g^{\mu \nu}$.

Next, we define the concept of an ${\sf R}$-tensor, which is important
when changing the CBF.  We consider a new CBF such that the
orientation matrix ${\sf R}'$ with respect to the new frame is related
to the old orientation ${\sf R}$ by

\begin{equation}
{\sf R}' =  {\sf R} {\sf U}(q),
\label{r69}
\end{equation}
where ${\sf U}(q) \in SO(3)$ is a smooth function of $q$.  The
coordinates $q^\mu$ are held fixed.  The quantities ${\bf
r}_\alpha$ and ${\sf S}_\alpha$ transform via

\begin{eqnarray}
{{\bf r}'}_\alpha & = & {\sf U}^T {\bf r}_\alpha, 
\label{r71} \\
{{\sf S}'}_\alpha & = & {\sf U}^T {\sf S}_\alpha,
\end{eqnarray}
where we have omitted the $q$ dependence.  We call ${\bf r}_\alpha$ a
rank one ${\sf R}$-tensor because it has one Latin index which
transforms with one copy of ${\sf U}^T$.  In general, an ${\sf
R}$-tensor transforms by contracting each Latin index with ${\sf U}^T$
as in Eq.~(\ref{r71}).  The rank of the ${\sf R}$-tensor is the number
of Latin indices it possesses.  Other rank one ${\sf R}$-tensors
include ${\bf n}_\alpha$, $\mbox{\boldmath $\omega$}_\alpha$, and ${\bf J}$.  Rank
zero ${\sf R}$-tensors, also called ${\sf R}$-scalars, do not depend
on the choice of CBF and include $T$ and $\dot{q}^\mu$.  Rank two
${\sf R}$-tensors include ${\sf M}$ and ${\sf M}_\alpha$.

The quantity ${\bf A}_\mu$ is not an ${\sf R}$-tensor.  Instead, it
has a more complicated transformation property

\begin{equation}
{{\bf A}'}_\mu = {\sf U}^T {\bf A}_\mu - \mbox{\boldmath $\gamma$}_\mu,
\end{equation}
where $\mbox{\boldmath $\gamma$}_\mu \times = {\sf U}^T {\sf U}_{, \mu}$.  An
interpretation of this transformation property as the gauge transformation law of
a non-Abelian gauge potential is given in Ref.~\cite{Littlejohn97}.  We
simply note that such an analysis motivates the introduction of the
Coriolis field strength,

\begin{equation}
{\bf B}_{\mu \nu} = {\bf A}_{\nu, \mu} -  {\bf A}_{\mu, \nu} 
- {\bf A}_\mu \times {\bf A}_\nu.
\label{r60}
\end{equation}
The Coriolis field strength is a rank two $q$-tensor and a rank
one ${\sf R}$-tensor.  We will only use the Coriolis field strength
briefly in Sect.~\ref{s9}.  However, it plays a central role in the
gauge theoretic approach to rovibrational coupling.

One should be aware that other quantities which have been introduced
are also not ${\sf R}$-tensors.  These include $h_{\mu \nu}$ and
$\tilde{\sf M}$.  However, it should be noted that both of these
quantities have counterparts, $g_{\mu \nu}$ and ${\sf M}$
respectively, which are ${\sf R}$-tensors.  For further insight into 
the relationship between these two pairs of quantities, see
Ref.~\cite{Littlejohn97}.  Various other quantities which are not
${\sf R}$-tensors include $\mbox{\boldmath $\tau$}_{\alpha \mu}$, ${\bf a}_\mu$,
${\bf K}$, $\mbox{\boldmath $\omega$}$, and $p_\mu$.

An important observation is that the two decompositions of the kinetic
energy, Eqs.~(\ref{r18}) and (\ref{r21}), differ in their
transformation properties.  Specifically, Eq.~(\ref{r18}) decomposes
the kinetic energy into two terms which are $q$- and ${\sf R}$-scalars.
Equation~(\ref{r21}), on the other hand, decomposes the kinetic energy
into two terms which are $q$-scalars but not ${\sf R}$-scalars.  Thus,
the latter decomposition is dependent on the choice of CBF, whereas
the former is not.  For this reason, we view Eq.~(\ref{r18}) as the
fundamental rovibrational decomposition of the kinetic energy.
However, depending on the CBF convention, the decomposition of
Eq.~(\ref{r21}) may very well be easier to compute.  (See
Sect.~\ref{s9}.)  We refer to Ref.~\cite{Littlejohn97} for further
discussion of these decompositions.

\section{General expressions for the quantum kinetic energy}
\label{s1}
\subsection{The unscaled kinetic energy}

\label{s7} 

We temporarily abandon the specific system of a molecular complex in
order to present expressions for the quantum kinetic energy of a
general system.  The results of this section are largely similar to
previous work of Nauts and Chapuisat\cite{Nauts85}, which in turn
relies on several earlier references.  We also note that Van
der Avoird {\it et al.} have employed a similar formalism for the
water trimer \cite{vanderAvoird96}.  Here we will summarise relevant
aspects of these results to fix notation and to lay the foundation for
Sect.~\ref{s8}.  Also, we apply the formalism to the simple example of
a single rigid body, which will be of future use.  In Sect.~\ref{s6} we
present a new approach to scaling the wave function.

We denote the classical kinetic energy by

\begin{equation}
T = \frac{1}{2} \pi_a G^{ab}(x) \pi_b,
\label{r27}
\end{equation}
where $x$ stands for a collection of generalised position variables
$x^a$, $a = 1,\ldots,d$ ($d$ is the number of degrees of freedom),
$\pi_a$, $a = 1,\ldots,d$, are generalised momenta, and $G^{ab}$ are
the components of the inverse metric tensor ${\sf G}^{-1}$.  When
repeated, the indices $a$, $b$, $c$, $\ldots$ are assumed to be summed
from $1$ to $d$, both in Eq.~(\ref{r27}) and the subsequent
development. The momenta $\pi_a$ are linear
combinations of the canonical  momenta,

\begin{equation}
\pi_a = C_a^{\; \; b}( x ) p_b,
\end{equation}
where $p_a$ is the momentum canonically conjugate to $x_a$ and the
$C_a^{\; \; b}$ are components of the change of basis matrix.  Nauts
and Chapuisat \cite{Nauts85} call the more general momenta $\pi_a$
quasi-momenta and reserve the term momenta for what we call the
canonical momenta $p_a$.  The kinetic energy Eq.~(\ref{r27}) is expressed in
terms of the canonical momenta $p_a$ by

\begin{equation}
T = \frac{1}{2} p_c  C_a^{\; \; c} G^{ab}  C_b^{\; \; d} p_d
= \frac{1}{2} p_a  \tilde{G}^{ab} p_b,
\end{equation}
where $\tilde{G}^{ab}$ are the components of the inverse metric
$\tilde{\sf G}^{-1} = {\sf C}^T {\sf G}^{-1} {\sf C}$ with respect to
the canonical momenta.  From now on, we omit the explicit $x$
dependence.

The quantum kinetic energy $\hat{T}$ is often expressed in the
Podolsky form \cite{Podolsky28}

\begin{equation}
\hat{T} 
=  \frac{1}{2} {1\over \sqrt{\tilde{G}}} 
\hat{p}_a \sqrt{\tilde{G}} \tilde{G}^{ab} 
\hat{p}_b, 
\label{r36}
\end{equation}
where $\tilde{G} = \det \tilde{\sf G}$ and where $\hat{p}_a$ is the momentum
operator conjugate to $x^a$,

\begin{equation}
\hat{p}_a =  -i {\partial \over \partial x^a},
\label{r35}
\end{equation}
setting $\hbar = 1$.  The operator $\hat{T}$ may also be expressed using
the adjoints of the momentum operators,

\begin{equation}
\hat{T} = \frac{1}{2} \hat{p}_a^\dagger  \tilde{G}^{ab} \hat{p}_b,
\label{r26}
\end{equation}
or more generally,

\begin{equation}
\hat{T} = \frac{1}{2} \hat{\pi}_a^\dagger G^{ab} \hat{\pi}_b,
\label{r25}
\end{equation}
where $\hat{\pi}_a$ is the operator corresponding to the classical
momentum $\pi_a$,

\begin{equation}
\hat{\pi}_a 
=  C_a^{\; \; b} \hat{p}_b.
\label{r29}
\end{equation}
As observed by van der Avoird {\it et al.} \cite{vanderAvoird96}, even
though the adjoint form of the kinetic energy represents the same
differential operator as the Podolsky form, Eqs.~(\ref{r26}) and
(\ref{r25}) are convenient for evaluating matrix elements since the
adjoint of the momentum operator effectively acts on the bra to the
left.

Equations~(\ref{r26}) and (\ref{r25}) are straightforward
consequences of the definition of the adjoint.  For an arbitrary
operator $\hat{A}$, the matrix element of $\hat{A}^\dagger$ with
respect to wave functions $\Phi$ and $\Phi'$ is

\begin{equation}
\left< \Phi \left| \hat{A}^\dagger \Phi' \right. \right> 
= \left< \hat{A} \Phi \left| \Phi' \right. \right>,
\label{r30}
\end{equation}
where the inner product is defined via

\begin{equation}
\left< \Phi \left| \Phi' \right. \right> 
= \int dv \; \Phi^* \Phi',
\label{r31}
\end{equation}
and the volume element is

\begin{equation}
dv = \sqrt{\tilde{G}} \, dx^1 \ldots dx^d 
= {\sqrt{G} \over |\det {\sf C}|} \, dx^1 \ldots dx^d, 
\label{r40}
\end{equation}
where $G = \det {\sf G}$.  From this definition, one finds that the momenta
$\hat{p}_a$ are not in general Hermitian but rather satisfy the
following relation

\begin{equation}
\hat{p}_a^\dagger = {1\over \sqrt{\tilde{G}}} \hat{p}_a \sqrt{\tilde{G}}.
\label{r37}
\end{equation}
More generally, the momenta  $\hat{\pi}_a$ satisfy

\begin{equation}
\hat{\pi}_a^\dagger 
= \hat{\pi}_a + {1 \over \sqrt{\tilde{G}}} 
\left[ \hat{p}_b  \sqrt{\tilde{G}} C_a^{\; \; b} \right],
\label{r33}
\end{equation}
where the square bracket notation indicates that $\hat{p}_b$ acts only
on the terms inside the brackets.  Equations~(\ref{r36}), (\ref{r37}),
and (\ref{r29}) combine to prove Eqs.~(\ref{r26}) and (\ref{r25}).

An important illustration of the preceding formalism and one which we shall need later is that of a single rigid
body.  We define Euler angles $[x^1, x^2, x^3] = [\alpha, \beta,
\gamma]$ in the usual way by ${\sf R}(\alpha,\beta,\gamma) = {\sf
R}_z(\alpha){\sf R}_y(\beta){\sf R}_z(\gamma)$, where ${\sf R}$
rotates the space frame into the body frame and ${\sf R}_i$ is a
rotation about the $i$th space axis.  The body referred angular
momenta $[\pi_1, \pi_2, \pi_3] = [J_1, J_2, J_3]$ are noncanonical
momenta related to the canonical momenta $[p_1,p_2,p_3] = [p_\alpha,
p_\beta, p_\gamma]$ via \cite{Biedenharn81a}

\begin{equation}
\left[
\begin{array}{c}
J_1 \\
J_2 \\
J_3
\end{array}
\right]
= {\sf C}
\left[
\begin{array}{c}
p_\alpha \\
p_\beta \\
p_\gamma
\end{array}
\right],
\end{equation}
where

\begin{equation}
{\sf C} =
\left[
\begin{array}{ccc}
-{\cos \gamma \over \sin \beta} & \sin \gamma & \cos \gamma \cot \beta \\
{\sin \gamma \over \sin \beta} & \cos \gamma & -\sin \gamma \cot \beta \\
0 & 0 & 1 
\end{array}
\right].
\label{r38}
\end{equation}
The classical kinetic energy in terms of the angular momenta is $T =
{\bf J} \cdot {\sf M}^{-1} {\bf J}/2$ where ${\sf M} = {\sf G}$ is the
body referred moment of inertia tensor, which is independent of the
Euler angles.  The volume element is readily computed from
Eq.~(\ref{r40}) to be

\begin{equation}
dv 
= \sqrt{\det {\sf M}} \sin \beta \, d\alpha \, d\beta \, d\gamma
= 8\pi^2 \sqrt{\det{\sf M}} \, d{\sf R}, 
\end{equation}
where $d{\sf R} = \sin \beta \, d\alpha \, d\beta \, d\gamma /(8 \pi^2)$ is the
normalised Haar measure on $SO(3)$.  The quantum kinetic energy is
expressed in terms of the operators $\hat{J}_i$

\begin{equation}
\left[
\begin{array}{c}
\hat{J}_1 \\
\hat{J}_2 \\
\hat{J}_3
\end{array}
\right]
= {\sf C}
\left[
\begin{array}{c}
-i \partial / \partial \alpha \\
-i \partial / \partial \beta \\
-i \partial / \partial \gamma
\end{array}
\right].
\label{r39}
\end{equation}
Using the volume element $dv$, one may verify that $\hat{J}_i$ is
Hermitian.  (On a deeper level, $\hat{J}_i$ is Hermitian because it is
a symmetry of the kinetic energy.)  Thus, the quantum kinetic energy
Eq.~(\ref{r25}) acquires the familiar form

\begin{equation}
\hat{T} = {1 \over 2} \hat{{\bf J}} \cdot {\sf M}^{-1} \hat{{\bf J}}.
\end{equation}

\subsection{The scaled kinetic energy}

\label{s6}

Often it is useful to multiply the original wave function $\Phi$ by some real positive
function ${\cal S}(x)$ to form a new wave function $\Psi$,

\begin{equation}
\Psi = {\cal S} \Phi.
\label{r28}
\end{equation}
Such a scaling produces a new kinetic energy operator acting on the
new wave function $\Psi$.  In this section, we derive the form of this
new kinetic energy operator.  Similar discussions are given by Nauts
and Chapuisat~\cite{Nauts85} and Chapuisat, Belafhal, and
Nauts~\cite{Chapuisat91}.  The most notable distinction between our
approach and these earlier accounts is our introduction of a new
adjoint, shown in Eq.~(\ref{r99}).  This adjoint allows for a
different form for the scaled kinetic energy operator shown in Eq.~(\ref{r101})
and the associated extrapotential term in Eq.~(\ref{r59}).

We note that Eq.~(\ref{r28}) induces a new inner product on the scaled
wave functions.  We denote this new inner product with an ${\cal S}$
subscript and define it via,

\begin{equation}
\left< \Psi | \Psi' \right>_{\cal S} 
= \left<  {1 \over {\cal S}} \Psi \left|  {1 \over {\cal S}} \Psi' \right. \right> 
= \left< \Psi \left|  {1 \over {\cal S}^2} \Psi' \right. \right> 
= \int {\sqrt{\tilde{G}} \over {\cal S}^2} dx^1 \ldots dx^d \; \Psi^* \Psi' .
\end{equation} 
Thus, $\sqrt{\tilde{G}} / {\cal S}^2 dx^1 \ldots dx^d$ is the volume
element associated with the scaled wave functions.  The operator
adjoint taken with respect to this new inner product will in general
be different from the adjoint taken with respect to the old inner
product.  To avoid confusion we will denote the new adjoint by
$\hat{A}^{\dagger({\cal S})}$.  These two adjoints are related by
the following computation

\begin{equation}
\left< \Psi \left| \hat{A}^{\dagger({\cal S})} \Psi' \right. \right>_{\cal S} 
= \left< \hat{A} \Psi \left| \Psi' \right. \right>_{\cal S} 
= \left< \hat{A} \Psi \left| {1 \over {\cal S}^2} \Psi' \right. \right> 
= \left< \Psi \left|  \hat{A}^\dagger {1 \over {\cal S}^2} \Psi' \right. \right> 
= \left< \Psi \left|  {\cal S}^2 \hat{A}^\dagger {1 \over {\cal S}^2} \Psi' \right. \right>_{\cal S}, 
\end{equation}
which summarises as

\begin{equation}
\hat{A}^{\dagger({\cal S})} = {\cal S}^2 \hat{A}^\dagger {1 \over {\cal S}^2}.
\label{r99}
\end{equation} 
Equation~(\ref{r99})  combines with Eq.~(\ref{r33}) to yield

\begin{equation}
\hat{\pi}^{\dagger ({\cal S})}_a
= \hat{\pi}^\dagger_a - 2 [ \hat{\pi}_a \ln {\cal S} ].
\label{r34}
\end{equation}
 
The scaling of the wave function transforms the kinetic energy
operator into $\hat{T}_{\cal S} = {\cal S} \hat{T} (1 / {\cal S})$.  Combining
Eqs.~(\ref{r25}) and (\ref{r99}), we find

\begin{equation}
\hat{T}_{\cal S} 
= \frac{1}{2} \left( {1 \over {\cal S}} \hat{\pi}^{\dagger ({\cal S})}_a {\cal S} \right)
G^{ab} \left( {\cal S}\hat{\pi}_b  {1 \over {\cal S}} \right).
\end{equation}
By more or less straightforward commutation of operators in the above
equation and using Eqs.~(\ref{r33}) and (\ref{r34}), we arrive at the
main result of this section

\begin{equation}
\hat{T}_{\cal S} 
= \frac{1}{2} \hat{\pi}_a^{\dagger ({\cal S})} G^{ab} \hat{\pi}_b + V_{\cal S},
\label{r101}
\end{equation}
where

\begin{eqnarray}
V_{\cal S} 
& = & -\frac{1}{2} \left( G^{ab} [ \hat{\pi}_a \ln {\cal S}] [\hat{\pi}_b \ln {\cal S}]
+ [ \hat{\pi}_a^{\dagger ({\cal S})} G^{ab} [ \hat{\pi}_b \ln {\cal S} ] ]\right) 
\nonumber \\
& = & \frac{1}{2} \left( G^{ab} [ \hat{\pi}_a \ln {\cal S}] [ \hat{\pi}_b \ln {\cal S}]
- [ \hat{\pi}_a^\dagger G^{ab} [ \hat{\pi}_b \ln {\cal S} ] ]\right). 
\label{r59}
\end{eqnarray}
Comparing Eq.~(\ref{r101}) to the unscaled expression Eq.~(\ref{r25}),
we note that the two operators differ by the additional scalar term in
Eq.~(\ref{r101}) and the different adjoints which are used.  Thus,
scaling the wave function may be used to place the adjoint of the
momenta in an alternative, perhaps more attractive, form, but only at
the expense of introducing an extrapotential term into the kinetic
energy.

\section{The quantum kinetic energy of a molecular complex}

\label{s8}

We quantise the classical kinetic energy Eq.~(\ref{r18}) using
Eq.~(\ref{r25}) derived in the previous section.  This approach
requires the operators $\hat{p}_\mu$, $\hat{J}_i$, and their adjoints.
Since the classical momentum $p_\mu$ is canonically conjugate to
$q^\mu$, the quantised operator $\hat{p}_\mu$ has the usual form of
Eq.~(\ref{r35})

\begin{equation}
\hat{p}_\mu = -i {\partial \over \partial q^\mu}.
\end{equation}
The quantised angular momenta $\hat{J}_i$ satisfy the standard
``anomalous'' commutation relations $[\hat{J}_i, \hat{J}_j] = -i
\sum_k \epsilon_{ijk} \hat{J}_k$ and of course commute with all rotationally
invariant operators, for example,

\begin{equation}
[ \hat{J}_i, {\sf M} ] 
= [ \hat{J}_i, \tilde{\sf M} ] 
= [ \hat{J}_i, {\bf A}_\mu ] 
= [ \hat{J}_i, {\bf a}_\mu ] 
= [ \hat{J}_i, g_{\mu \nu} ] 
= [ \hat{J}_i, h_{\mu \nu} ] 
= [ \hat{J}_i, \hat{p}_\mu ] 
= 0.
\label{r41}
\end{equation}

To compute the volume element of Eq.~(\ref{r40}), we require explicit
coordinates covering all directions of configuration space.  This
means defining three Euler angles $[\theta^1,\theta^2,\theta^3 ] =
[\alpha, \beta, \gamma]$, describing the collective orientation ${\sf
R}$, which complement the $d - 3$ internal coordinates $q^\mu$.  Here,
$d = 3n + 3n_n + 2n_c -3$ is the dimension of the centre of mass
system.  We adopt the Euler angle conventions used in
Section~\ref{s1}.  The noncanonical momenta $\hat{\pi}_i = \hat{J}_i$
and $\hat{\pi}_\mu = \hat{p}_\mu - \hat{\bf J} \cdot {\bf A}_\mu$ are
expressed in terms of the canonical momenta $\hat{p}_i = -i \partial /
\partial \theta^i$ and $\hat{p}_\mu$ by

\begin{equation}
\left[
\begin{array}{c}
\hat{J}_i \\
\hat{p}_\mu - {\hat{\bf J}} \cdot {\bf A}_\mu 
\end{array}
\right]
=
\left[
\begin{array}{cc}
C_i^{\; \;j} & 0 \\
- \sum_k A^k_\mu C_k^{\; \; j} & \delta_\mu^{\; \; \nu}
\end{array}
\right]
\left[
\begin{array}{c}
\hat{p}_j \\
\hat{p}_\nu 
\end{array}
\right],
\label{r42}
\end{equation}
where $C_i^{\;\; j}$ are the components of the matrix in
Eq.~(\ref{r38}) and the sum over $j$ is implicit.  The full $d \times
d$ matrix in Eq.~(\ref{r42}) corresponds to the matrix $C_a^{\;\; b}$
in Eq.~(\ref{r29}).  Its determinant is equal to the determinant of
the upper left block alone, that is $\det {\sf C} = -1/\sin \beta$.
The metric $G_{ab}$ with respect to the momenta $\pi_i = J_i$ and
$\pi_\mu = p_\mu - {\bf J} \cdot {\bf A}_\mu$ is seen from
Eq.~(\ref{r18}) to have determinant

\begin{equation}
G  = g \det {\sf M},
\label{r70}
\end{equation}
where $g = \det g_{\mu \nu}$.  The volume element is therefore

\begin{eqnarray}
dv 
& = & {\sqrt{ G }\over |\det{\sf C}| } 
\, d\alpha \, d\beta \, d\gamma \, dq^1 \ldots dq^{d-3}
=  \sqrt{g \det {\sf M}} \sin \beta 
\, d\alpha \, d\beta \, d\gamma \, dq^1 \ldots dq^{d-3} 
\nonumber \\
& = & 8\pi^2 \sqrt{g \det {\sf M}} 
\, d{\sf R} \, dq^1 \ldots dq^{d-3}.
\end{eqnarray}
An identity we will use later is the following alternative expression
for $G$

\begin{equation}
G = h \det \tilde{\sf M},
\label{r66}
\end{equation}
where $h$ is the determinant of $h_{\mu \nu}$.  This identity follows
from the fact that the change of basis connecting Eq.~(\ref{r18}) with
Eq.~(\ref{r21}) is orthogonal.

Since $\det{\sf M}$ and $g$ are rotationally invariant, their presence
in $dv$ is irrelevant for the computation of $\hat{J}_i^\dagger$.
Therefore, the computation of $\hat{J}_i^\dagger$ reduces to the case
of a single rigid rotor examined in Section~\ref{s1} from which we
recall that $\hat{J}_i^\dagger = \hat{J}_i$.  Therefore, from
Eqs.~(\ref{r18}) and (\ref{r25}), we find

\begin{equation}
\hat{T} = {1\over 2} \hat{\bf J} \cdot {\sf M}^{-1}  \hat{\bf J} +
{1\over 2} (\hat{p}_\mu^\dagger - {\bf A}_\mu \cdot \hat{\bf J})  
 g^{\mu \nu} (\hat{p}_\nu - {\bf A}_\nu \cdot \hat{\bf J}).
\label{r61}
\end{equation}
The ordering of the operators $\hat{J}_i$ with respect to the other
factors is irrelevant, on account of Eq.~(\ref{r41}).  The ordering of
the $\hat{p}_\mu$ with respect to $g^{\mu \nu}$ and ${\bf A}_\mu$,
however, is essential.  Note that the $\hat{p}_\mu$ are not in general
Hermitian but rather satisfy

\begin{equation}
\hat{p}_\mu^\dagger 
= {1 \over \sqrt{G}} \hat{p}_\mu \sqrt{G}
= {1\over \sqrt{g \det {\sf M} }} \hat{p}_\mu \sqrt{g \det {\sf M}},
\label{r57}
\end{equation}
as easily seen from Eq.~(\ref{r37}) and the fact that $\tilde{G} = G / (\det {\sf
C})^2 = G (\sin \beta)^2$.

We now scale the wave function by a factor

\begin{equation}
{\cal S} = \sqrt{8 \pi^2} G^{1/4} =  \sqrt{8 \pi^2} (\det {\sf M})^{1/4} g^{1/4}
\end{equation}
to obtain a new form of the kinetic energy.  First, we note that the
transformed volume element is

\begin{equation}
{dv \over {\cal S}^2} 
= d{\sf R} \, dq^1 \ldots dq^{d-3}.
\end{equation}
The angular momenta $\hat{J}_i$ are still Hermitian with respect to
this new volume element, that is $\hat{J}_i^{\dagger({\cal S})} =
\hat{J}_i$, as may be noted from Eq.~(\ref{r34}) and the fact that
$\cal S$ is rotationally invariant.  However, since the new volume
element contains no $q^\mu$ dependence in the Jacobian prefactor, we
have the added benefit that $\hat{p}_\mu$ is now Hermitian, that is

\begin{equation}
\hat{p}_\mu^{\dagger({\cal S})} = \hat{p}_\mu.
\end{equation}
Therefore, the transformed kinetic energy of Eq.~(\ref{r101}) takes the simple form

\begin{equation}
\hat{T}_{\cal S} = {1\over 2} \hat{\bf J} \cdot {\sf M}^{-1}  \hat{\bf J} +
{1\over 2} (\hat{p}_\mu - {\bf A}_\mu \cdot \hat{\bf J})  
 g^{\mu \nu} (\hat{p}_\nu - {\bf A}_\nu \cdot \hat{\bf J}) + V_{\cal S},
\end{equation}
where the extrapotential term may be reduced to 

\begin{equation}
V_{\cal S}  = {1 \over 2} G^{-1 /4 } \left[ {\partial \over \partial q^\mu} 
g^{\mu \nu}
{\partial \over \partial q^\nu} G^{1 / 4}   \right].
\label{r58}
\end{equation}
We observe that $V_{\cal S}$ is an ${\sf R}$-scalar, but not a
$q$-scalar.  Therefore, $V_{\cal S}$ depends on the choice of internal
coordinates, but not on the choice of CBF.  Further discussion of this
matter is given in Ref.~\cite{Littlejohn97}.

The quantum kinetic energy may also be placed in a form
analogous to Eq.~(\ref{r21}).  The unscaled kinetic energy
Eq.~(\ref{r61}) becomes 

\begin{equation}
\hat{T} = {1\over 2} (\hat{\bf J} - \hat{\bf K}^\dagger) 
\cdot \tilde{\sf M}^{-1} (\hat{\bf J} - \hat{\bf K}) 
+ {1\over 2} \hat{p}^\dagger_\mu  h^{\mu \nu} \hat{p}_\nu,
\end{equation}
where 

\begin{equation}
\hat{\bf K} = h^{\mu \nu} {\bf a}_\mu \hat{p}_\nu.
\end{equation}
Similarly, the scaled kinetic energy becomes
 
\begin{equation}
\hat{T}_{\cal S} =  {1\over 2} (\hat{\bf J} - \hat{\bf K}^{\dagger ({\cal S})}) 
\cdot \tilde{\sf M}^{-1} (\hat{\bf J} - \hat{\bf K}) 
+ {1\over 2} \hat{p}_\mu  h^{\mu \nu} \hat{p}_\nu + V_{\cal S}.
\end{equation}

\section{Example: a monomer-atom complex}

\label{s9}

We compute the rovibrational kinetic energy explicitly for a system
containing a single noncollinear rigid monomer with moment of inertia
${\sf M}_1$ and a single atom, for example, $\mbox{Ar-NH}_3$.  The
kinetic energy of such systems has already been studied by Brocks and
van Koeven\cite{Brocks88}, van der Avoird \cite{vanderAvoird93}, and
Makarewicz and Bauder \cite{Makarewicz95}.  Our presentation is mainly
designed to illustrate the formalism of the preceding sections,
although we believe that the derivation of the Coriolis potential
$A^i_\mu$, Coriolis field strength $B^i_{\mu \nu}$, and internal
metric $g_{\mu \nu}$ is new.

We define the CBF by fixing it to the rigid monomer.  This implies that
the matrix ${\sf S}$, defining the orientation of the monomer's IBF in
the CBF, is constant.  We take this constant to be the identity,

\begin{equation}
{\sf S}(q) = {\sf I}.
\label{r46}
\end{equation}
Since there is only one rigid body and one Jacobi vector, we drop all
``$\alpha$'' subscripts, except on ${\sf M}_1$, where the ``$1$'' serves
to distinguish the moment of inertia of the monomer from the total
moment of inertia ${\sf M}$ of the complex.  The Jacobi vector ${\bf
r}$ locates the atom with respect to the monomer and its components
may therefore be chosen as the internal coordinates, that is

\begin{equation}
[q^1,q^2, q^3] = [r_1, r_2, r_3].
\end{equation}
Thus,

\begin{equation}
r_{i,\mu} = \delta_{i \mu}.
\label{r47}
\end{equation}
Furthermore, from Eqs.~(\ref{r16}) and (\ref{r46}), we have

\begin{equation}
\mbox{\boldmath $\tau$}_\mu = 0.
\label{r48}
\end{equation}
Inserting Eqs.~(\ref{r46}), (\ref{r47}), and (\ref{r48}) into  
Eqs.~(\ref{r63}) -- (\ref{r65}), we readily obtain

\begin{eqnarray}
{\sf M} & = &  
r^2 {\sf I} - {\bf r} {\bf r}^T + {\sf M}^{i}_1, 
\label{r49} \\
a_{i \mu} & = & ({\bf r} \times {\bf r}_{,\mu})_i = \sum_j \epsilon_{ij\mu} r_j, 
\label{r50} \\
h_{\mu \nu} & = & {\bf r}_{, \mu} \cdot {\bf r}_{, \nu} = \delta_{\mu \nu},
\label{r51}
\end{eqnarray}
where the $i$ superscript on ${\sf M}^{i}_1$ indicates that it is
referred to the monomer's IBF (which agrees here with the CBF) and is
hence a constant matrix.  

We proceed to first construct the kinetic energy Eq.~(\ref{r21}) which
results here in a simpler form than Eq.~(\ref{r18}).  Using
Eqs.~(\ref{r19}) and (\ref{r62}) we compute

\begin{eqnarray}
\tilde{\sf M} & = & {\sf M}^{i}_1, \\
{\bf K} & = & {\bf r} \times {\bf p},
\end{eqnarray}
where ${\bf p} = [p_1, p_2, p_3]$.  These are particularly simple results and together with
Eq.~(\ref{r51}) yield the classical kinetic energy

\begin{equation}
T = \frac{1}{2} ({\bf J} -  {\bf r} \times {\bf p}) \cdot  ({\sf M}^{i}_1)^{-1}
({\bf J} -  {\bf r} \times {\bf p}) + \frac{1}{2} {\bf p} \cdot {\bf p}.
\label{r64}
\end{equation}
The quantum kinetic energy requires the further result

\begin{equation}
G = h \det \tilde{\sf M} = \det {\sf M}^{i}_1,
\end{equation}
which follows from Eq.~(\ref{r66}) and shows that $G$ is constant.
Hence, from Eq.~(\ref{r57}) it is clear that $\hat{p}_\mu$ is
Hermitian with respect to the original inner product.  Since $\hat{\bf
K} = {\bf r} \times \hat{\bf p}$ we find that $\hat{\bf K}$ is also
Hermitian with respect to both the original and the scaled inner
products. Furthermore, the extrapotential term of Eq.~(\ref{r58})
arising in the scaled kinetic energy vanishes.  Thus, both the
original and the scaled quantum kinetic energies are identical and
each is formed by simply replacing ${\bf p}$ and ${\bf J}$ in
Eq.~(\ref{r64}) with $\hat{\bf p}$ and $\hat{\bf J}$ respectively.
Our results agree with earlier derivations by Brocks and van Koeven
\cite{Brocks88} and van der Avoird\cite{vanderAvoird93}.

To simplify the algebra in constructing the kinetic energy
Eq.~(\ref{r18}), we assume the rigid body is a spherical top with
${\sf M}_1^{i} = \kappa {\sf I}$.  The total moment of inertia
tensor given in Eq.~(\ref{r49}) may be explicitly inverted, with the
form

\begin{equation}
{\sf M}^{-1} 
= {1 \over r^2 + \kappa} {\sf I} 
+ {1 \over \kappa (r^2 + \kappa)} {\bf r} {\bf r}^T,
\label{r53}
\end{equation}
and combined with Eqs.~(\ref{r52}) and (\ref{r20}) to yield 
explicit forms for the Coriolis potential and the internal metric,

\begin{eqnarray}
A^i_{\mu } & = & \sum_j {1 \over r^2 + \kappa } \epsilon_{ij\mu} r_j, 
\label{r54} \\
g_{\mu \nu} & = & {1 \over r^2 + \kappa} ( \kappa \delta_{\mu \nu} + r_\mu r_\nu).
\label{r56}
\end{eqnarray}
The inverse of the internal metric is 

\begin{equation}
g^{\mu \nu}  =  {1 \over \kappa} [ (r^2 + \kappa)  \delta_{\mu \nu} - r_\mu r_\nu ]. 
\label{r55} 
\end{equation}
Eqs.~(\ref{r53}), (\ref{r54}), and (\ref{r55}) combine with
Eq.~(\ref{r18}) to yield an explicit form for the classical kinetic
energy.  As earlier, the quantum kinetic energy, both original and
scaled, is obtained by simply replacing ${\bf p}$ and ${\bf J}$ by
their operator counterparts, without the need for Hermitian
conjugates or an extrapotential term.

It is interesting to compute the Coriolis field strength defined by
Eq.~(\ref{r60}),

\begin{equation}
B_{\mu \nu}^i = \epsilon_{\mu \nu \sigma}{1 \over (r^2 + \kappa)^2}
(r_i r_\sigma + 2 \kappa \delta_{i \sigma}).
\end{equation}
As the separation $r$ of the atom from the monomer goes to infinity,
the Coriolis field strength tends toward $B^i_{\mu \nu} \rightarrow
\epsilon_{\mu \nu \sigma} r_i r_\sigma/r^4$.  We change the CBF, as in
Eq.~(\ref{r69}), via a matrix ${\sf U}({\bf r})$ which rotates
$\hat{\bf z}$ into ${\bf r}/r$.  Then, since ${\bf B}_{\mu \nu}$ is a
rank one ${\sf R}$-tensor, as $r$ goes to infinity, the new field
strength tensor approaches

\begin{equation}
{\bf B}_{\mu \nu}' = {\sf U}^T {\bf B}_{\mu \nu}
\rightarrow \epsilon_{\mu \nu \sigma} {r_\sigma \over r^3} \hat{\bf z}.
\end{equation}
The above asymptotic form is that of a (non-Abelian) monopole field
\cite{Goddard77}.  A similar monopole field is already known to exist
in the three-body problem \cite{Iwai87a,Iwai87b}, a fact which
has led to several useful applications~\cite{Iwai87a,Iwai87b,Montgomery96,Mitchell97}.  We remark that the above
asymptotic form is valid even if the monomer is an asymmetric top.

\section{Conclusions}

\label{s10}

We have computed the kinetic energy of an arbitrary molecular complex
for arbitrary coordinate and body frame conventions.  In so doing, we
have tried to provide an efficient framework in which explicit
Hamiltonians may be readily computed for specific choices of
coordinates and frames.  We have provided a discussion of
transformation properties to facilitate the changing of these
conventions.  Our formalism is illustrated with the example of a
monomer-atom system, and more complex systems may be handled with
similar ease within our framework.

One of the more novel and intriguing aspects of our derivation is the
appearance of the Coriolis potential and the various insights which
are possible by adopting a gauge theoretical viewpoint.  We briefly
cite two areas of current research which are based on this
perspective.  First, using gauge theoretic reasoning we have managed
to generalise the Eckart conditions, so often employed for small
vibrations in molecules, to systems of rigid bodies.  Much of the
formalism for small vibrational analysis in molecules can then be
readily ported over to study small amplitude vibrations in clusters of
rigid molecules.  Second, we have been able to understand rotational
splittings in molecules with internal rotors as a sort of Coriolis
Aharonov-Bohm effect.  These applications and others will be the
subject of future publications.

\section{Acknowledgements}
The authors gratefully acknowledge stimulating discussions with
Professor R. J. Saykally and the astute review of the manuscript by
Dr. M. M\"uller.  This work was supported by the U. S. Department of
Energy under Contract No. DE-AC03-76SF00098.

\end{document}